# Orbital Arrangements and Magnetic Interactions in the Quasi-One-Dimensional Cuprates $A$CuMoO$_4$(OH) ($A$ = Na, K)

Kazuhiro Nawa*, Takeshi Yajima, Yoshihiko Okamoto†, and Zenji Hiroi

Institute for Solid State Physics, The University of Tokyo, Kashiwa, Chiba 277-8581, Japan

**ABSTRACT:** A new spin-1/2 quasi-one-dimensional antiferromagnet KCuMoO$_4$(OH) is prepared by the hydrothermal method. The crystal structures of KCuMoO$_4$(OH) and the already-known Na-analogue, NaCuMoO$_4$(OH), are isotypic, comprising chains of Cu$^{2+}$ ions in edge-sharing CuO$_4$(OH)$_2$ octahedra. Despite the structural similarity, their magnetic properties are quite different because of the different arrangements of $d_{x^2-y^2}$ orbitals carrying spins. For NaCuMoO$_4$(OH), $d_{x^2-y^2}$ orbitals are linked by superexchange couplings via two bridging oxide ions, which gives a ferromagnetic nearest-neighbor interaction $J_1$ of −51 K and an antiferromagnetic next-nearest-neighbor interaction $J_2$ of 36 K in the chain. In contrast, a staggered $d_{x^2-y^2}$ orbital arrangement in KCuMoO$_4$(OH) results in superexchange couplings via only one bridging oxide ion, which makes $J_1$ antiferromagnetic as large as 238 K and $J_2$ negligible. This comparison between the two isotypic compounds demonstrates an important role of orbital arrangements in determining the magnetic properties of cuprates.

## Introduction

Quantum spin systems attract much attention because quantum fluctuations may suppress conventional magnetic order and can induce exotic states such as a spin liquid.[1,2] The most promising candidates are found in compounds that contain Cu$^{2+}$ ions with spin 1/2. The Cu$^{2+}$ ion has a $d^9$ electronic configuration with one unpaired electron occupying either the $d_{x^2-y^2}$ or $d_{z^2}$ orbital in the octahedral crystal field. This degeneracy is eventually lifted by lowering symmetry due to the Jahn–Teller effect: when two opposite ligands move away and the other four ligands come closer, the $d_{x^2-y^2}$ orbital is selected to carry the spin, whereas the $d_{z^2}$ orbital is selected in the opposite case. Since this Jahn–Teller energy is quite large, either distortion of octahedra always occurs in actual compounds. Since magnetic interactions between neighboring Cu spins are caused by superexchange interactions via overlapping between Cu $d$ orbitals and O ligand $p$ orbitals, the arrangements of the Cu $d$ orbitals and their connections via ligands are important to determine the magnetic properties of cuprates.

Among many cuprates, a rich variety of crystal structures are found in natural minerals and their synthetic analogues. Recently, for example, many copper minerals with Cu$^{2+}$ ions in the kagome geometry have been focused on from the viewpoint of quantum magnetism: Volborthite (Cu$_3$V$_2$O$_7$(OH)$_2$•2H$_2$O),[3-5] Herbertsmithite (Zn$_{1-x}$Cu$_{3+x}$(OH)$_6$Cl$_2$),[6] Vesignieite (BaCu$_3$V$_2$O$_8$(OH)$_2$),[7] and so on. Besides, there are a lot of minerals where Cu$^{2+}$ forms one-dimensional (1D) chains, such as Chalcanthite (CuSO$_4$•5H$_2$O),[8] Linarite (PbCuSO$_4$(OH)$_2$),[9] and Natrochalcite (NaCu$_2$(SO$_4$)$_2$(OH)•H$_2$O).[10] Natrochalcite also has many synthetic analogues of $A$Cu$_2$(SO$_4$)$_2$(OH)•H$_2$O ($A$ = K, Rb, Ag) and $A$Cu$_2$(SeO$_4$)$_2$(OH)•H$_2$O ($A$ = Na, K, Rb, Ag, Tl, NH$_4$).[11,12]

In the present study, we focus on two cuprates which crystallize in the Adelite (CaMgAsO$_4$(OH))–Descloizite (PbZnVO$_4$(OH)) structure type with the general formula $ATX$O$_4$(OH)[13] ($A$ = Na, Ca, Cd, Pb, … ; $T$ = Mn, Fe, Cu, Zn ,… ; $X$ = V, As, Si, Mo, … ). This structure contains $T$O$_4$(OH)$_2$ octahedra that form chains by edge sharing, and the chains are separated by $X$O$_4$ tetrahedra and $A$ atoms. Thus, it is expected that their magnetic properties have a 1D character. Although many compounds are known to crystallize in this structure, their magnetic properties have not yet been studied in detail except for NaCuMoO$_4$(OH).[14] NaCuMoO$_4$(OH) was first synthesized hydrothermally by Moini et al. in 1986.[15] It crystallizes in an orthorhombic structure with the space group $Pnma$. The Cu$^{2+}$ atom occupies the $T$ site to form a chain made of CuO$_4$(OH)$_2$ octahedra. Since the CuO$_4$(OH)$_2$ octahedron is heavily elongated perpendicular to the edge-sharing plaquette of the ligands, the $d_{x^2-y^2}$ orbital lying in the plaquette is selected to carry spin 1/2. Then, magnetic interactions between neighboring Cu spins occur by superexchange interactions via two bridging oxide ions. Since the Cu–O–Cu bond angles in this type of chains are approximately 90°, the nearest-neighbor interaction $J_1$ is expected to be ferromagnetic according to the Goodenough–Kanamori rule. In addition, there is a super-superexchange interaction $J_2$ through Cu–O–O–Cu paths along the chain, which is antiferromagnetic. In fact, the exchange interactions of NaCuMoO$_4$(OH) are estimated as $J_1$ = −51 K and $J_2$ = 36 K.[14] Note that the two interactions cause frustration in the spin chain, because $J_1$ favors all the spins in the same direction whereas $J_2$ tends to make every second spins antiparallel. This frustration can lead to a variety of magnetic phases such as spin nematic order, which is theoretically predicted to be a sort of spin liquid with an unusual spin order analogous to the nematic state in liquid crystals[16-19] but has not yet been observed in actual compounds[20].

In this study, we have searched for a new compound in this series of compounds and succeeded to synthesize a K analogue, KCuMoO$_4$(OH). Its single crystals were grown by the hydrothermal method, and the crystal structure and magnetic properties are investigated. The crystal structures of KCuMoO$_4$(OH) and NaCuMoO$_4$(OH) are isotypic, but, surprisingly, their magnetic properties are totally different. It is shown that the difference comes from a subtle difference in the atomic coordinates, which causes different arrangements of $d_{x^2-y^2}$ orbitals.

### Crystal Growths and Characterizations

Single crystals of KCuMoO$_4$(OH) were synthesized by the hydrothermal method. First, 0.4619 g (1.912 mmol) of Cu(NO$_3$)$_2$·3H$_2$O was dissolved in 10 ml of water. Then, 9.525 g (40 mmol) of K$_2$MoO$_4$ was added and the solution was stirred until it became homogeneous. The solution was put in a Teflon beaker of 25 ml volume, placed in a stainless steel autoclave, heated to 433 K, and then slowly to 493 K in 240 hours. Greenish crystals with the maximum size of 0.6×0.6×0.6 mm$^3$ were obtained, as shown in Fig. 1(a). The crystals were characterized by means of powder X-ray diffraction (XRD) using Cu $K_\alpha$ radiation (Rigaku, RINT-2000), chemical composition analysis using inductively coupled plasma (ICP) spectroscopy (Horiba, JY138KH), and thermal gravimetry (Bruker AXS TG-DTA2020SAH). A powder XRD pattern from crushed crystals (Fig. S1 in supplemental information) shows that all the diffraction peaks are indexed based on an orthorhombic structure with the space group *Pnma,* indicating that there is no impurity inclusion. The obtained lattice constants of $a$ = 7.9901(1) Å, $b$ = 6.3201(1) Å, and $c$ = 9.5818(1) Å are slightly larger than those of the Na compound ($a$ = 7.726 Å, $b$ = 5.968 Å, and $c$ = 9.495 Å[15]), reflecting the larger ionic radius of the K$^+$ ion compared with the Na$^+$ ion. The chemical compositions of K, Mo, and Cu determined by ICP are 13.5(1), 22.0(1), and 34.0(2) wt%, respectively, which are close to the stoichiometric compositions expected for KCuMoO$_4$(OH): 14.0, 22.7, and 34.3 wt%. A dehydration reaction was observed upon heating above 623 K with a weight loss of 3.2(1)%, which indicates that a half mole of H$_2$O per formula unit has been lost as expected from the stoichiometric composition. Therefore, we have successfully obtained a new compound KCuMoO$_4$(OH), which is obviously related to the Na analogue.

### Crystal Structure

To determine the crystal structure, XRD data on a single crystal were collected in a R-AXIS RAPID IP diffractometer with a monochromated Mo-$K_\alpha$ radiation ($\lambda$ = 0.71075 Å) at 296 K. The structure was refined against |F$^2$| by the direct method using the SHELXL2013 software. A multi-scan absorption correction was employed, and the displacement parameters were refined as anisotropic. The H atom could not be located. Structural refinements have converged with a reasonably small *R* factor of 2.04%: the crystal data and atomic parameters are listed in Tables 1 and 2, respectively. Further details of the refinement parameter, selected bond lengths, and bond valence sums can be obtained from Table S1, S2, and S3, respectively, which are included in the supporting information. Figure 1(b) shows the crystal structure of the K compound. Basically, it is identical to that of the Na compound, in which CuO$_4$(OH)$_2$ octahedra form edge-sharing chains along the $b$ axis, and the chains are separated by the MoO$_4$ tetrahedra and K(Na) atoms. However, because of the size difference between K$^+$ and Na$^+$ ions, the shapes of the edge-sharing chains are slightly but significantly different between the two compounds, as will be mentioned later.

### Magnetic Properties

Despite the structural similarities between the two compounds, their magnetic properties are quite different. Figure 2 shows the temperature dependences of magnetic susceptibility $\chi$ measured at a magnetic field of 1 T. The $\chi$ of the Na compound increases with decreasing temperature, following a Curie–Weiss (CW) law with a CW temperature of -5.0 K. Then, it exhibits a broad peak at 14 K due to an antiferromagnetic short-range order.[14] In sharp contrast, for the K compound, the magnitude of $\chi$ is much smaller and the peak temperature lies at a much higher temperature of 150 K. These indicate that there are stronger antiferromagnetic interactions in the K compound. In fact, the CW temperature is estimated to be -205 K for the $\chi$ in the range of 250–300 K at $H$ // $a$. Note that this CW temperature can be an underestimate because of the limited fitting range which is not high enough compared with the peak temperature. Both the magnitude and the peak temperature of the data in the wide temperature range of 80-300 K are well reproduced by a spin-1/2 1D antiferromagnetic chain model,[21] in which the $g$ factor and the nearest-neighbor interaction $J_1$ are used as variables together with a temperature-independent contribution $\chi_0$. The fitting shown by the dashed curve in Fig. 2 for $H \parallel a$ is reasonably good and yields $g = 1.92(1)$, $J_1 = 196(1)$ K, and $\chi_0 = 1.19(5) \times 10^{-4}$ cm$^3$ mol$^{-1}$. Thus, a simple spin chain with a large antiferromagnetic interaction may be realized in the K compound, in contrast to the frustrated $J_1$–$J_2$ chain with $J_1 = -51$ K and $J_2 = 36$ K in the Na compound.[14] In addition, the $\chi$ of the K compound exhibits a large anisotropy below the peak temperature, while the Na compound has little anisotropy in $\chi$; there is a small anisotropy arising from that of the g factor in the Na compound. The $\chi$ at $H \parallel b$ is enormously enhanced toward $T = 0$, while those at $H \parallel a$ and $c$ decrease, followed by small upturns below ~10 K. Note that the $\chi$ value at 2 K for $H \parallel b$ becomes 8 times as large as those for $H \parallel a$ and $c$. These features are sample-independent, suggesting that the origin is not extrinsic from defects or magnetic impurities but intrinsic, which will be discussed later.

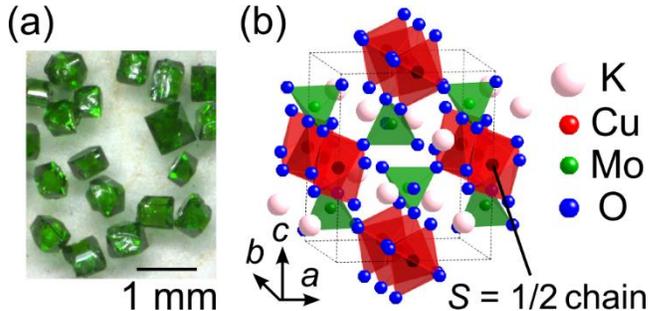

Figure 1. (a) Photograph of single crystals of KCuMoO$_4$(OH). (b) Crystal structure of KCuMoO$_4$(OH) (hydrogen atoms are omitted). CuO$_4$(OH)$_2$ octahedra share their edges to form a chain of $S$ = 1/2 along the $b$ axis.

Table 1. Crystal data for KCuMoO$_4$(OH).

| Crystal Data | |
| --- | --- |
| Formula | KCuMoO$_4$(OH) |
| Crystal size | 0.10×0.08×0.05 mm$^3$ |
| Crystal system | Orthorhombic |
| Space group | *Pnma* |
| $a$ (Å) | 7.98385(17) |
| $b$ (Å) | 6.31560(14) |
| $c$ (Å) | 9.5641(4) |
| $V$ (Å$^3$) | 482.25(3) |
| $Z$ | 4 |
| Calculated Density (g cm$^{-3}$) | 3.851 |

**Table 2.** Atomic positions and displacement parameters for KCuMoO$_4$(OH)

| Atom | site | x | y | z | $U_{11}$ | $U_{22}$ | $U_{33}$ | $U_{12}$ | $U_{23}$ | $U_{13}$ |
|---|---|---|---|---|---|---|---|---|---|---|
| Cu | 4b | 0.5 | 0 | 0 | 0.0098(3) | 0.0072(3) | 0.0088(3) | 0.00187(14) | -0.00132(13) | -0.00047(15) |
| K | 4c | 0.37445(9) | 0.25 | 0.33160(10) | 0.0132(4) | 0.0189(4) | 0.0121(4) | 0 | 0.0019(3) | 0 |
| Mo | 4c | 0.11152(3) | -0.25 | 0.20063(3) | 0.00605(18) | 0.00793(18) | 0.0083(2) | 0 | 0.00100(9) | 0 |
| O(1) | 4c | 0.2919(3) | -0.25 | 0.0942(3) | 0.0117(12) | 0.0176(12) | 0.0139(13) | 0 | 0.0029(10) | 0 |
| O(2) | 8d | 0.1039(2) | 0.5159(3) | 0.30492(19) | 0.0139(9) | 0.0104(9) | 0.0123(10) | -0.0008(6) | 0.0028(6) | -0.0001(7) |
| O(3) | 4c | -0.0594(3) | -0.25 | 0.0869(3) | 0.0106(12) | 0.0343(16) | 0.0173(14) | 0 | -0.0044(11) | 0 |
| O(4) | 4c | 0.3792(3) | 0.25 | 0.0538(3) | 0.0089(11) | 0.0095(12) | 0.0104(13) | 0 | -0.0009(8) | 0 |

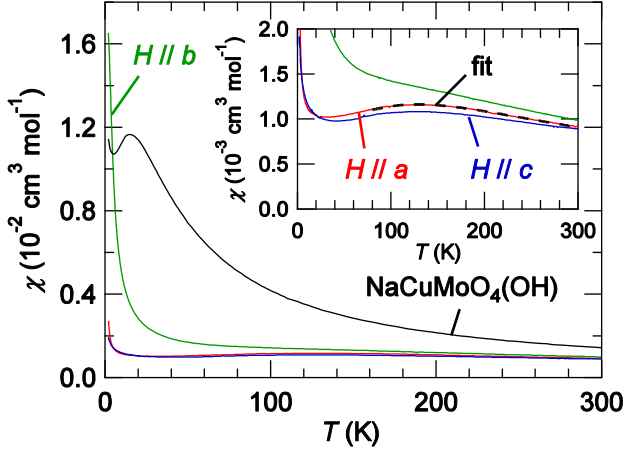

Figure 2. Temperature dependences of magnetic susceptibility measured in a magnetic field of 1 T upon heating after zero-field cooling. The red, green, and blue curves are obtained at $H \parallel a$, $b$, and $c$, respectively, for KCuMoO$_4$(OH). The black curve represents the $\chi$ of NaCuMoO$_4$(OH) for comparison[14]; $\chi$ of the polycrystalline sample is shown since its anisotropy is small. The inset shows a blowup around the broad maximum. The dashed curve represents a fit to a 1D antiferromagnetic chain model[21].

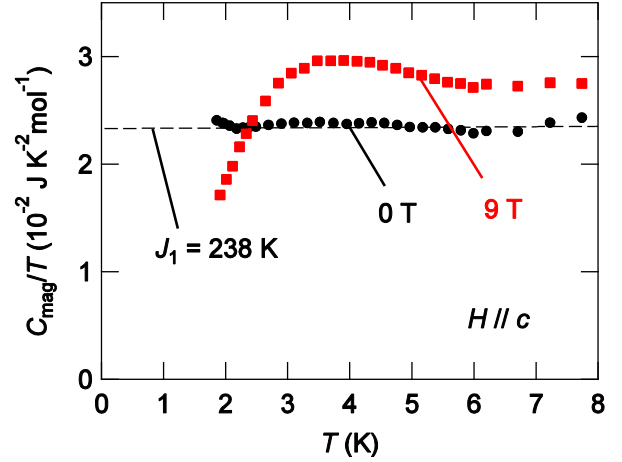

Figure 3. Temperature dependences of magnetic heat capacity divided by temperature in magnetic fields of 0 and 9 T. The dashed line on the zero-field data represents a fit to a 1D antiferromagnetic chain with $J_1 = 238$ K.[21]

Heat capacity measurements also confirm that KCuMoO$_4$(OH) is a spin-1/2 1D antiferromagnet. Figure 3 shows a magnetic heat capacity divided by temperature, $C_{mag}/T$, obtained by subtracting a lattice contribution estimated by the Debye model from the raw data. The $C_{mag}/T$ measured at zero field does not show any anomaly down to 2 K. Thus, the transition temperature should be lower than 2 K, which is two orders of magnitude smaller than the CW temperature, indicating a good one-dimensionality in magnetic interactions. Note that the $C_{mag}/T$ becomes almost constant at low temperatures below ~6 K. Such a $T$-linear dependence in $C_{mag}$ is not due to electron heat capacity, since this compound is insulating. Alternatively, the $T$-linear heat capacity is characteristic of a spin-1/2 1D antiferromagnetic chain, and is given by $C_{mag} = 2R/3J_1 \times T$, where $R$ is the gas constant.[21, 22] The observed $C_{mag}/T = 0.0234(1)$ J K$^{-2}$ mol$^{-1}$ yields $J_1 = 238(1)$ K, which is in reasonably good agreement with $J_1 = 196(1)$ K from the $\chi$ data. We take $J_1 = 238$ K from the heat capacity data more reliable, because there are some ambiguities in fitting the $\chi$ data. On the other hand, a marked change in $C_{mag}$ is observed when a magnetic field is switched on. Applying 9 T field enhances the $C_{mag}$ above 2.5 K and reduces below 2.5 K, indicating that a significant amount of magnetic entropy has been pushed to higher temperatures. This suggests that an energy gap opens by the magnetic field, as discussed later.

**Discussions**

It is now clear that the two compounds NaCuMoO$_4$(OH) and KCuMoO$_4$(OH) are isotypic in the crystal structure but show very different magnetic properties. Here we discuss on the magnetic interactions in them and also the unusual magnetic susceptibility of the K compound. As schematically illustrated in Fig. 4(a), the Na compound has two competing intrachain interactions of $J_1 = -51$ K and $J_2 = 36$ K, while the K compound has a much stronger antiferromagnetic interaction of $J_1 = 238$ K. This striking difference must be originated from local structures. Let us compare the shapes and linkages of CuO$_4$(OH)$_2$ octahedra between the two compounds in Fig. 4(b); note that all octahedra are crystallographically identical in each compound. There are 3 oxygen sites forming the octahedron: O1 and O4 (probably OH) bridge two neighboring octahedra, and O2 is connected to only one Cu. In the Na compound, the octahedron is elongated toward O2 with a large difference in bond lengths: $d$(Cu–O1) = 2.075 Å and $d$(Cu–O4) = 1.898 Å are much shorter than $d$(Cu–O2) = 2.386 Å. Thus, it is no doubt that the $d_{x^2-y^2}$ orbital extending toward O1 and O4 is responsible for $S = 1/2$. On the other hand, in the K compound, the octahedron is elongated toward O1: $d$(Cu–O2) = 2.044 Å and $d$(Cu–O4) = 1.920 Å are much shorter than $d$(Cu–O1) = 2.463 Å. This geometry makes a spin-carrying $d_{x^2-y^2}$ orbital extend toward O2 and O4. Therefore, the CuO$_4$(OH)$_2$ octahedra are distorted in different ways in the two compounds, resulting in different arrangements of $d_{x^2-y^2}$ orbitals.

To be emphasized here is that magnetic couplings between Cu spins should critically depend on the arrangements of $d_{x^2-y^2}$ orbitals. $J_1$ occurs by superexchange interactions via two paths, Cu–O1–Cu and Cu–O4–Cu in the Na compound, while via only one path Cu–O4–Cu in the K compound. These interactions must be sensitive to the bond angles. According to the Goodenough–Kanamori rule, $J_1$ becomes ferromagnetic when a Cu–O–Cu bond angle is close to 90° and changes into antiferromagnetic when the bond angle exceeds 95–98°.[23,24] In the Na compound, the Cu–O1–Cu path with a bond angle of 91.97° must be ferromagnetic, while the Cu–O4–Cu path with a bond angle of 103.65° may be antiferromagnetic: ferromagnetic $J_1$ may be attained from the dominant contribution of the former path. In contrast, in the K compound, the Cu–O4–Cu path with a large bond angle of 110.60° must make $J_1$ strongly antiferromagnetic. On the other hand, $J_2$ is dominated by super-superexchange interactions via Cu–O1–O4–Cu and Cu–O4–O1–Cu paths, which actually exist in the $d_{x^2-y^2}$ arrangement of the Na compound but not in that of the K compound; little overlap between the $d_{x^2-y^2}$ and the $p$ orbitals of O1 makes $J_2$ negligible. Therefore, the difference in magnetic interactions in the two compounds is well understood in terms of the local structures.

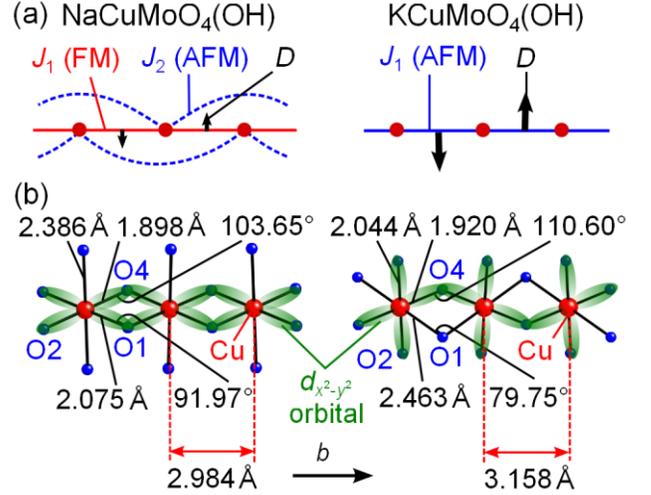

Figure 4. (a) Schematic representations of magnetic interactions present in the chains of NaCuMoO$_4$(OH) and KCuMoO$_4$(OH). $J_1$ and $J_2$ denote nearest-neighbor and next-nearest neighbor interactions, respectively, and $D$ refers to Dzyaloshinskii-Moriya vector. Abbreviations FM and AFM represent ferromagnetic and antiferromagnetic interactions, respectively. (b) Local structures of the chains made of CuO$_4$(OH)$_2$ octahedra and the arrangements of $d_{x^2-y^2}$ orbitals (green lobes).

The different local structures in the two compounds are apparently attributed to the size difference of the Na$^+$ and K$^+$ ions. These ions are located between two Cu chains (Fig. 1(b)) and determine some important bond lengths in the chains. Particularly, the Cu–Cu distances are significantly different: $d$(Cu–Cu) is 2.984 and 3.158 Å in the Na and K compounds, as shown in Fig. 4(b). The former is close to those in other compounds having edge-sharing chains where $d_{x^2-y^2}$ orbitals are linked by two oxide atoms as in the Na compound: $d$(Cu–Cu) = 2.905 and 2.941 Å for LiVCuO$_4$[25] and CuGeO$_3$,[26] respectively. In contrast, the latter long Cu–Cu distance may not be compatible with this type of chains but prefer a different local structure with a large Cu–O–Cu bond angle, as realized in the K compound. Therefore, even in the isotypic structures, the two types of local structures and thus of orbital arrangements are made possible.

Next, we will consider the origin of the unusual behavior in the magnetic susceptibility of KCuMoO$_4$(OH); the large enhancement at low temperatures only for $H // b$. Similar behavior has been observed in Cu-benzoate,[27–29] Yb$_4$As$_3$,[30] and KCuGaF$_6$[31] and well explained by the presence of staggered Dzyaloshinskii-Moriya (DM) interactions.[32,33] We think that a similar mechanism applies to the present compound. The model Hamiltonian considered is given as

$$H = \sum_i \{J_1 \boldsymbol{S}_i \cdot \boldsymbol{S}_{i+1} - g\mu_B H S_{iz} + (-1)^i \boldsymbol{D} \cdot (\boldsymbol{S}_i \times \boldsymbol{S}_{i+1})\}, (1)$$

where $\mu_B$, $H$, and $D$ denote the Bohr magneton, magnetic field, and the DM vector, respectively. The DM interactions tend to modify a simple antiferromagnetic alignment of spins on a chain; they serve to incline spin $S_{i+1}$ with respect to spin $S_i$. When all DM vectors are identical, spin $S_{i+1}$ is always inclined in the same direction with respect to spin $S_i$ so as not to generate uncompensated magnetic moments in an antiferromagnetic chain. However, when DM vectors are staggered in orientation as depicted in Fig. 4(a), the direction of this inclination alternates so as to generate net moments. In other words, staggered DM interactions in an antiferromagnetic chain classically stabilize two-sublattice canted antiferromagnetism. This leads to a large increase in $\chi$ at low temperatures below a broad maximum, where an antiferromagnetic short-range order develops, even in a paramagnetic state as observed for the K compound. Moreover, this enhancement should occur effectively when $H$ is perpendicular to $D$ and be absent for $H // D$, because spins are inclined to directions perpendicular to magnetic field for $H // D$. The observed anisotropic behavior for the K compound, a large enhancement in $\chi$ only for $H // b$, is understood if the $D$ is perpendicular to the $b$ axis.

Generally speaking, a possible direction of $D$ is restricted by crystal symmetry. In the Na and K compounds, there is a mirror plane that bisects the Cu–Cu bond and a twofold screw axis that passes through the Cu sites. The former requires $D$ to align perpendicular to the chain ($D_b$ = 0), and the latter requires $D$ to alternate along the chain direction ($D_{i,a} = -D_{i+1,a}$, $D_{i,c} = -D_{i+1,c}$). Thus, the DM interactions are completely staggered with the DM vectors always perpendicular to the $b$ axis in both compounds, as depicted in Fig. 4(a). Concerning the magnitude of $D$, one expects that it is larger in the K compound because the alternation of $d_{x^2-y^2}$ orbital along the chain is absent in the Na compound but present in the K compound. This is the reason why an enhancement of $\chi$ is prominent in the K compound.

It is also characteristic for a spin chain with staggered DM interactions that an energy gap is induced by an applied magnetic field. In fact, such a gap opening has been extensively studied in Cu-benzoate through heat capacity[29], inelastic neutron scattering[28, 29] and ESR experiments[34]. Thus, the field-induced energy gap observed in the heat capacity data of the K compound also supports the presence of staggered DM interactions. More detail analyses on the magnetic properties will be given elsewhere.

The dramatic interplay between crystal structures and magnetic interactions observed in NaCuMoO$_4$(OH) and KCuMoO$_4$(OH) is interesting and unique. A similar observation is found in related compounds such as CuCl$_2$,[35–38] CuCl$_2$•2H$_2$O[37, 39, 40] and CuCl$_2$•2NC$_5$H$_5$,[41, 42] which contains spin chains of edge-sharing CuCl$_4$X$_2$ ($X$ = Cl, O, and N, respectively) octahedra. In these compounds, the different ligands may induce changes in the orbital arrangement. On one hand, "orbital switching" from the $d_{z^2}$ to $d_{x^2-y^2}$ orbital observed in Volborthite, which is probably caused by a change in the hydrogen bonding of water molecules[4, 5], gives rise to a substantial change in magnetic interactions. However, these phenomena are essentially related to the changes of crystal symmetry. In sharp contrast, in the present compounds, only the small differences of atomic coordinates under the same crystal symmetry lead to completely different $d$ orbital arrangements and thus cause the large changes in the sign and magnitude of magnetic interactions. In future work, we would like to synthesize solid solutions between the two compounds to control the magnetic interactions systematically.

## Conclusions

We have prepared single crystals of a new quasi-1D cuprate KCuMoO$_4$(OH) and investigated its crystal structure and magnetic properties in comparison with those of the Na analogue NaCuMoO$_4$(OH). The two compounds are structurally isotypic but have different arrangements of $d_{x^2-y^2}$ orbitals. As the result, the magnetic properties become critically different: NaCuMoO$_4$(OH) provides a frustrated $J_1$–$J_2$ chain system, while KCuMoO$_4$(OH) is a 1D antiferromagnet with $J_1$ = 238 K with staggered DM interactions. An interesting relationship between crystal structures and magnetic interactions is demonstrated by the comparison of the two compounds.

## Acknowledgement

We thank M. Koike and M. Isobe for chemical analyses.

## Supporting Information

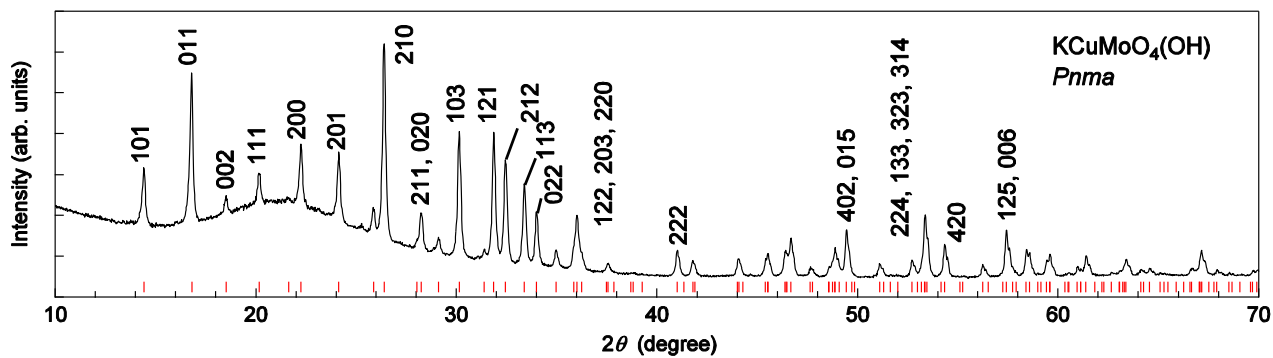

Figure S1. Powder XRD pattern of crushed crystals of KCuMoO$_4$(OH) measured by the Debye–Scherrer method. The vertical red lines at the bottom indicate the calculated positions of reflections. Miller indices of selected reflections are given above their peaks. The broad background around 20° is from a glass capillary tube.

Table S1. Refinement parameters for KCuMoO$_4$(OH)

| Data collection and Refinement | |
| --- | --- |
| Diffractometer | Rigaku R-AXIS RAPID |
| Radiation type | Mo $K_\alpha$ (0.71075 Å) |
| $2\theta_{max}$ (°) | 54.938 |
| Number of reflections | 4489 |
| Unique reflections | 608 |
| Absorption correction | Multi-scan |
| $T_{min}$ | 0.443 |
| $T_{max}$ | 0.677 |
| reflections for refinement | 608 |
| $hkl$ limits | $-10 \leq h \leq 10$, $-7 \leq k \leq 8$, $-12 \leq l \leq 12$ |
| $\mu$ (mm$^{-1}$) | 7.788 |
| number of parameters | 46 |
| $R$ | 2.04% |
| $wR$ | 5.31% |
| $\Delta\rho_{max}$ (e Å$^{-3}$) | 0.78 |
| $\Delta\rho_{min}$ (e Å$^{-3}$) | -1.29 |

Table S3. Bond valence sums of cations

| Compound | Na/K | Cu | Mo |
| --- | --- | --- | --- |
| NaCuMoO$_4$(OH) | 0.960 | 2.089 | 5.862 |
| KCuMoO$_4$(OH) | 1.232 | 2.027 | 5.807 |

Table S2. Bond lengths (Å) for KCuMoO$_4$(OH)

| Bonds | Length | Bonds | Length |
| --- | --- | --- | --- |
| Cu–Cu | 3.15780(8) | K–O2' | 2.8074(19) (×2) |
| Cu–O1 | 2.463(2) (×2) | K–O3 | 2.853(3) |
| Cu–O2 | 2.0443(18) (×2) | K–O4 | 2.657(3) |
| Cu–O4 | 1.9204(14) (×2) | Mo–O1 | 1.764(3) |
| K–O1 | 2.841(3) | Mo–O2 | 1.784(2) (×2) |
| K–O2 | 2.7481(19) (×2) | Mo–O3 | 1.745(3) |